# Temperature and gate effects on contact resistance and mobility in graphene transistors by TLM and Y-function methods

Francesca Urban[1], Grzegorz Lupina[2], Alessandro Grillo[1], Nadia Martucciello[3], and Antonio Di Bartolomeo[1]*

[1] Physics Department, University of Salerno and CNR-Spin, 84084 Fisciano, Salerno, Italy; furban@unisa.it (F.U)

[2] IHP-Microelectronics, Im Technologiepark 25, 15236 Frankfurt (Oder), Germany; (currently with Infineon Technologies Austria)

[3] CNR-Spin, 84084 Fisciano, Salerno, Italy

* Correspondence: adibartolomeo@e-mail.com

**Abstract:** The metal-graphene contact resistance is one of the major limiting factors toward the technological exploitation of graphene in electronic devices and sensors. A high contact resistance can be detrimental to device performance and spoil the intrinsic great properties of graphene. In this paper, we fabricate graphene field-effect transistors with different geometries to study the contact and channel resistance as well as the carrier mobility as a function of gate voltage and temperature. We apply the transfer length method and the y-function method showing that the two approaches can complement each other to evaluate the contact resistance and prevent artifacts in the estimation of the gate-voltage dependence of the carrier mobility. We find that the gate voltage modulates the contact and the channel resistance in a similar way but does not change the carrier mobility. We also show that the raising temperature lowers the carrier mobility, has negligible effect on the contact resistance, and can induce a transition from a semiconducting to a metallic behavior of the graphene sheet resistance, depending on the applied gate voltage. Finally we show that eliminating the detrimental effects of the contact resistance on the transistor channel current almost doubles the carrier field-effect mobility and that a competitive contact resistance as low as 700 $\Omega \cdot \mu m$ can be achieved by the zig-zag shaping of the Ni contact.

# 1. Introduction

The isolation of graphene [1–3] in 2004 and, later on, of h-BN[4], phosphorene [16, 17], $MoS_2$ [6–13], $WSe_2$ [14–16], $PdSe_2$ [17,18], etc., has strongly attracted the interest of the material science community to the world of two-dimensional (2D) materials.

Graphene, the bidimensional layer of carbon atoms arranged in a honeycomb lattice, is still one of the most studied 2D systems for the unmatched electron mobility, the remarkable current capability and thermal conduction, the relatively high optical absorption coefficient, the mechanical strength as well as the easy and low cost fabrication [19–24].

Graphene is commonly produced by exfoliation from graphite [25,26], epitaxial growth on SiC [27] or chemical vapor deposition (CVD) [28,29]. In particular, CVD produces uniform and large-scale graphene flakes of high-quality and is compatible with the silicon technology; therefore, it has been largely exploited to realize new electronic devices such as diodes [30–33], transistors [34–36], field emitters [37,38], chemical-biological sensors [39,40], optoelectronic systems [41], photodetectors [42–47] and solar cells [48].

Due to its gapless band-structure, with the valence and conduction bands touching each other at the so-called Dirac points, graphene originates ambipolar field-effect transistors with V-shaped transfer characteristics, dominated by a p-branch at negative and n-type conduction at positive gate voltage [49]. The ambipolar conduction can be an important feature for complementary logic applications; however, the limited on/off ratio caused by the absence of intrinsic bandgap is a significant obstacle and requires delicate material engineering for real applications [50–52].

Despite the several doping techniques available to tune the graphene conductivity and boost the performance of graphene transistors, a major problem remains the suppression of device on-current caused by the graphene/contact resistance [53]. Indeed, ohmic and low resistance contacts are important figures of merit for high frequency devices and the realization of stable and low-resistance contacts is still under intensive study [54–58]. The variation of the contact resistance, $R_C$, is attributed to many different causes, related to graphene growth and number of layers, metal type and deposition process, quality of the metal graphene/interface, measurement conditions, etc.

Conventional ohmic contacts between graphene and various metals exhibit rather large contact resistance ranging from few hundreds to thousands $\Omega \cdot \mu m$. Studies have been conducted on various types of metal/graphene interfaces showing that the best contact resistances can be achieved with Ni and Cu contacts yielding $R_C$ as low as $\sim 300 \, \Omega$ [59–62]. Although the choice of the metal type is an important ingredient for good quality contacts, recent researches have developed special techniques for the reduction of the contact resistance. The most successful strategies have been the modification of the contact area to increase charge injection through the graphene edges and the graphene etching under the contacts to favor the formation of dangling carbon-to-metal bonds. Contact resistances down to $100 \, \Omega \cdot \mu m$ have been obtained in this way [60,62–64]. As reported by Anzi et al. [64] the under contact graphene etching reduces the contact resistance both for Au and Ni/Au contacts. Same results have been obtained by Smith [60] exploiting cut graphene under the contact edge, with a contact resistance dependence on the number of cuts. Lisker et al. [54] obtained interesting results on devices similar to the one presented in this report, showing that the increment of the contact perimeter favors the reduction of metal/graphene interface resistance, optimizing the contact on the graphene sheet.

A low contact resistance enables the study of intrinsic graphene properties and increases the performance of graphene devices. As matter of fact, the contact resistance can be tuned by the application of a gate voltage ($V_{gs}$), which modulates the doping of graphene. In this scenario, the contact resistance becomes larger in correspondence of the Dirac points, where the conductivity of graphene is suppressed [65,66].

The temperature dependence of $R_C$ in graphene devices is still a controversial topic. A conspicuous number of studies report discrepant results evidencing either a negligible dependence of $R_C$ on T or strong changes of contact resistance with temperature [67–69].

In this work, we investigate the effect of gate voltage and temperature on the contact and channel resistance and on the carrier mobility in graphene field-effect transistors with Ni contacts. We fabricate back-gated devices with multiple leads which we analyze by both the transfer length method (TLM) [70–74] and the Y-function method [75–77]. The complementary application of the two methods leads to a more robust estimation of the contact resistance and of the gate-voltage dependence of the carrier mobility. We show that the gate voltage modulates the contact and the channel resistance in a similar way but has negligible effect on the carrier mobility. We also find that the field effect mobility decreases with the raising temperature, which does not affect the contact resistance, but can induce a transition from a semiconducting to a metallic behavior in the channel resistance, depending on the gate voltage. Finally, we show that eliminating the detrimental effect of the contact resistance can result in more than 80% increase of the field effect mobility.

## 2. Materials and Methods

Graphene synthesis has been performed on Ge/Si substrates using Aixtron's Black Magic BM300T CVD tool. The synthesis was carried out at the deposition temperatures of $885°C$ using $CH_4$ as source of carbon and $Ar/H_2$ mixture as carrier gas. The pressure was kept at 700 mbar during the 60 min deposition in order to optimize the fabrication process [54,78]. The so-obtained graphene was transferred on p-type doped Si (5-20 $\Omega \cdot$ cm) capped with 100 nm $SiO_2$ layer. The graphene monolayer [54] was patterned in long stripes by electron beam lithography (EBL) and dry etching, then covered by PMMA to prevent damages. After that, Ni metal contacts were deposited by EBL, thermal metal evaporation and lift-off process. Different layouts were defined, with an example reported in Figure 1. The devices consist of patterned graphene stripes contacted with several parallel leads, at gradually increasing distances ($d_i$). Structures with diverse combinations of the contact size, distance and/or shape, were fabricated and analyzed as well.

Measurements at different temperatures were performed using a Janis probe station equipped with four metallic nanotips connected to the source-measurement units of a Keithley 4200 SCS (Tektronix Inc.), at pressure of $\sim 0.8$ mbar. The metal contacts were used as the drain and source electrodes while the $Si/SiO_2$ substrate as the back gate and the gate dielectric, respectively.

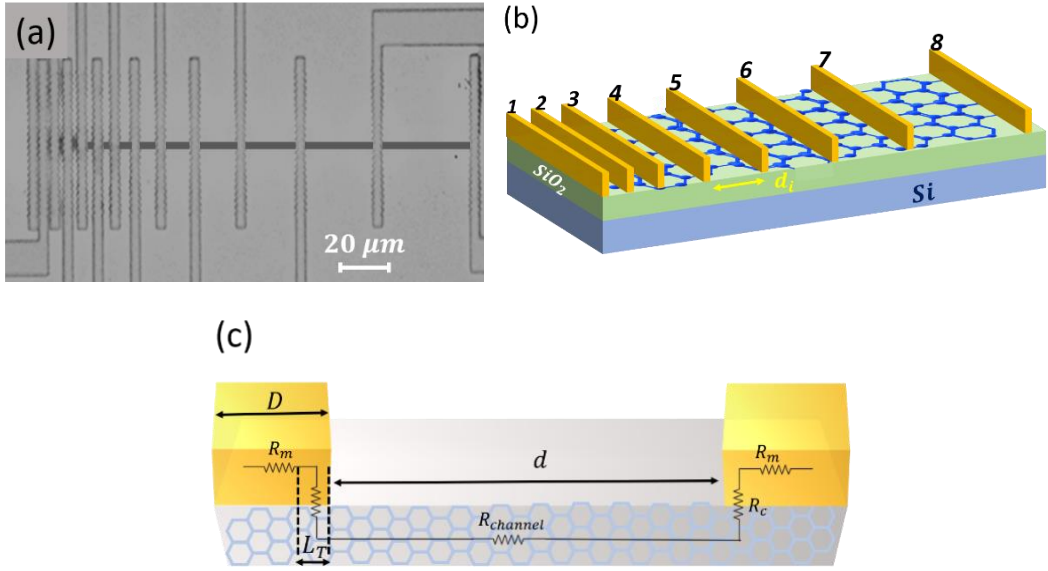

**Fig. 1.** (a) Optical image and (b) schematic of a device with zig-zag shaped contacts. (c) Schematic of the device, showing the metal ($R_m$), the contact ($R_c$) and the channel ($R_{channel}$) resistances. $L_T$ is the transfer length, representing the distance over which most of the current ($1 - 1/e$) flows between the contact and the channel.

The schematic of Figure 1(c) shows that the total resistance, $R_T$, obtained from the $I_{ds} - V_{ds}$ (drain-to-source current versus drain-to-source voltage) curves measured in a two-probe configuration between two given contacts, includes the contributions of the metal resistance, $R_m$, the contact resistance, $R_C$, i.e. the resistance at the 3D-metal/2D-graphene interface, and the channel resistance, $R_{channel}$:

$$R_T = R_{channel} + 2R_m + 2R_C \tag{1}$$

The channel resistance can be written as

$$R_{channel} = R_{sheet} \frac{d}{W} \tag{2}$$

where $R_{sheet}$ is the graphene sheet resistance in $\Omega/sq$ ($sq$ = square), W the width of the graphene stripe and $d$ the distance between the two chosen contacts. The metal resistance $R_m$ is orders of magnitude lower than $R_{channel}$ and can be neglected. The contact resistance, which can be comparable or higher than $R_{channel}$, can be expressed in terms of the transfer length, $L_T$, that represents the distance over which most of the current ($1 - 1/e$) flows between the contact and the channel: $R_C = R_{sheet} \frac{L_T}{W}$ [79]. Therefore

$$R_T = R_{sheet} \frac{d}{W} + 2R_C = \frac{R_{sheet}}{W}(d + 2L_T) \tag{3}$$

Eq. (3) is used to estimate $R_{sheet}$ and $R_C$ from the straight-line fitting of a $R_T$ vs $d$ plot (TLM plot). The intercept of the straight-line with the horizontal axis ($-2L_T$) provides the transfer length. If $L_T$ is small compared to the size D of the contact, the current flows mostly through the edge of the contact (current crowding effect) and only the contact edge influences the carrier injection and the conduction in the graphene channel. In this scenario, there are only two possibilities to reduce the contact resistance: etching the graphene under the contact to increase the contact edges or increasing the perimeter of the edge, for instance using zig-zag shaped edges.

For two-probe configuration measurements, an alternative approach to estimate the contact resistance is the so-called Y-function method (YFM). The method includes the contact resistance in the expression of the transistor current $I_{ds}$ as a function of $V_{ds}$ and $V_{gs}$ (the gate-to-source voltage) [76,80]:

$$I_{ds} = \frac{\frac{W}{L} \mu C_{ox} V_{ds}(V_{gs} - V_{Dirac})}{1 + \frac{W}{L} \mu C_{ox} R_C (V_{gs} - V_{Dirac})} \quad (4)$$

where $C_{ox}$ is the SiO$_2$ capacitance ($C_{ox} = 33\ nFcm^{-2}$ for 100 nm SiO$_2$), $\mu$ is the field-effect mobility and $V_{Dirac}$ is the gate voltage corresponding to the Dirac point, i.e. to the minimum of the $I_{ds} - V_{gs}$ characteristic of the graphene transistor.

The $V_{gs}$ derivative of eq. (4) is the transconductance

$$g_m = \frac{dI_{ds}}{dV_{gs}} = \frac{\frac{W}{L} \mu C_{ox} V_{ds}}{[1 + \frac{W}{L} \mu C_{ox} R_C (V_{gs} - V_{Dirac})]^2} \quad (5)$$

and the ratio

$$\frac{I_{ds}}{\sqrt{g_m}} = \sqrt{\frac{W}{L} \mu C_{ox} V_{ds}}\ (V_{gs} - V_{Dirac}) = Y \quad (6)$$

is the so-called Y-function. Y results independent of $R_C$, while $g_m^{-1/2}$ depends linearly on $(V_{gs} - V_{Dirac})$ with angular coefficient proportional to the contact resistance. Thus, the plots of $I_{ds} g_m^{-1/2}$ and $g_m^{-1/2}$ vs $(V_{gs} - V_{Dirac})$ can be exploited to obtain the mobility $\mu$ (which is not affected by the contact resistance) and the contact resistance $R_C$, respectively.

Finally, taking the derivative of the Y-function in eq. (6) with respect to $V_{gs} - V_{Dirac}$, we obtain the mobility unaffected by the contact resistance, which should be independent of $V_{gs} - V_{Dirac}$:

$$\mu = \left[\frac{dY}{d(V_{gs} - V_{Dirac})}\right]^2 \frac{L}{W C_{ox} V_{ds}} \quad (7)$$

## 3. Results

A two-probe configuration is adopted to measure the transfer ($I_{ds} - V_{gs}$ curve at given $V_{ds}$) and output ($I_{ds} - V_{ds}$ curves at selected $V_{gs}$) characteristics for different contact combinations. In Figure 2, we report an example of such measurements for the 2 $\mu m$ wide and 1 $\mu m$ long graphene channel contacted with Ni leads of size $D \sim 3\ \mu m$ and zig-zag shaped edges (see Figure 1(a)).

The transfer characteristic (Figure 2(a)) displays an asymmetric ambipolar behavior with a dominant hole branch and a current minimum (Dirac point) slightly above $V_{gs} = 0\ V$. The different slope of the two branches corresponds to the hole mobility ($\sim 150\ cm^2V^{-1}s^{-1}$) higher than the electron one ($\sim 100\ cm^2V^{-1}s^{-1}$), while the Dirac point at positive $V_{gs}$ indicates a low p-type doping due to adsorbates and process residues such as PMMA [56,81]. The hole-electron asymmetry is due to both unbalanced carrier injection from metal contacts and graphene interaction with the $SiO_2$ dielectric [81–86]. The interaction with $SiO_2$ is also the main cause of the hysteresis which appears when the gate voltage is swept back and forth [85–87]. The low mobility is otherwise attributed to the fabrication process which still yields graphene of limited quality and needs further optimization. Figure 2(b) shows a linear $I_{ds} - V_{ds}$ behavior confirming the ohmic nature of the Ni/graphene contacts.

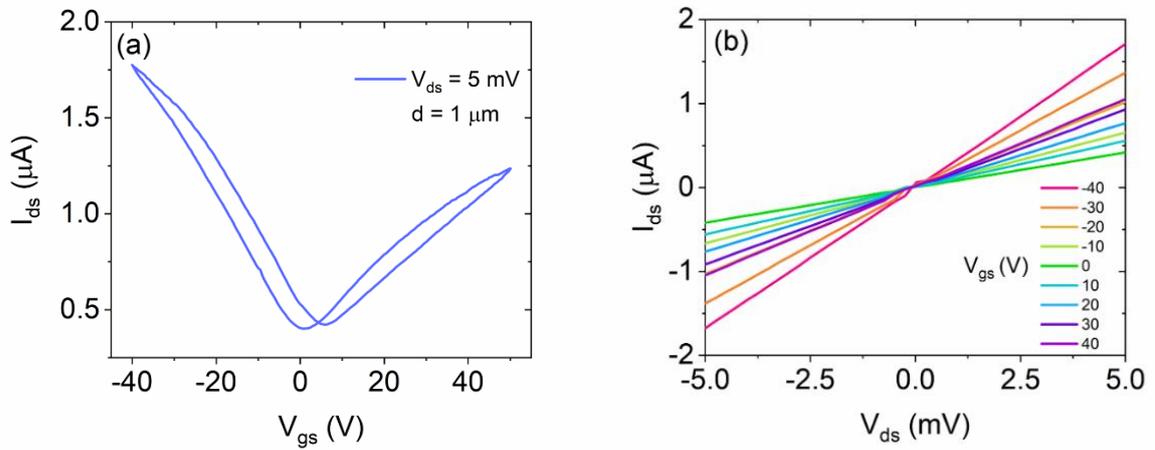

**Fig. 2.** (a) Transfer and (b) output characteristics of a graphene transistor with 2 $\mu m$ channel width and 1 $\mu m$ channel length.

Figure 3(a) shows the total resistance $R_T$ measured between multiple couples of leads of the TLM structure, at room temperature and under different gate biases, ranging from -40 V to 40 V. The TLM curves display the linear behavior predicted by eq. (3) and are used to extract $R_C$ and $R_{sheet}$ as a function of the gate voltage $V_{gs}$ (Figure 3(b)). Both parameters exhibit a non-monotonic trend with the maximum values ($R_C \sim 2.5\ k\Omega$ and $R_{sheet} \sim 14\ k\Omega/sq$) corresponding to the Dirac point ($V_{Dirac} \sim 0\ V$), and a decrease when the back-gate dopes the graphene by attracting electrons or holes in the channel. We highlight that Figure 3(b) demonstrates that the gate voltage affects the graphene layer not only in the channel region but also under the contacts, as previously reported [88].

$L_T$ extracted from the $R_T$ vs $d$ plot ranges between 300 $nm$ and 500 $nm$, which is small compared to the 3 $\mu m$ contact size, thereby confirming that the device works under the aforementioned current crowding regime.

Remarkably, comparison with a similar device contacted by straight contacts, i.e. no zig-zag edges, measured in the same conditions, exhibits ∼ 400% higher contact resistance (∼ 3.5 kΩ · μm vs ∼ 700 Ω · μm at $V_{gs} = -40V$), confirming the importance of increasing the length of the contact perimeter.

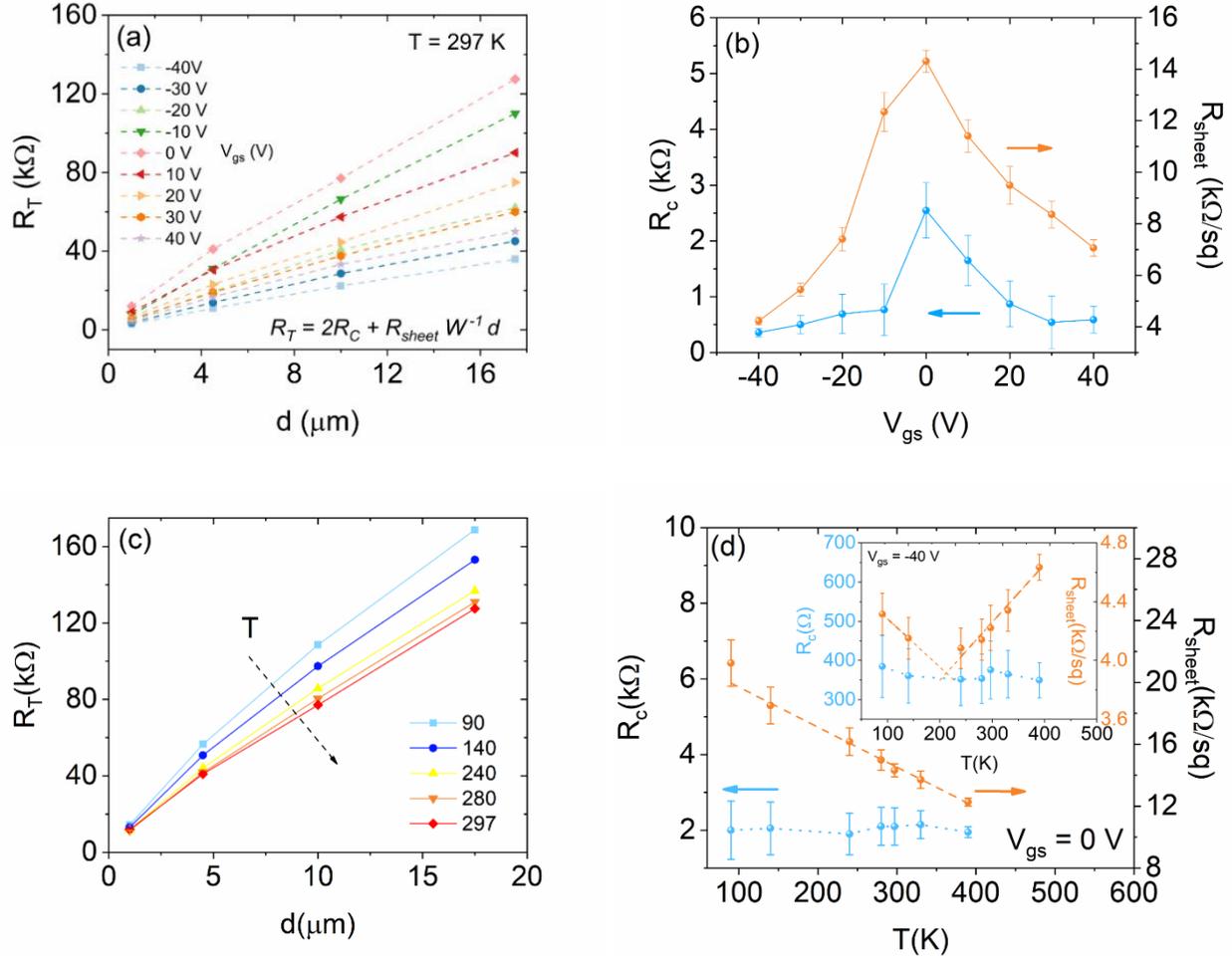

**Fig. 3**. (a) TLM plot and (b) $R_C$ and $R_{sheet}$ as function of the gate voltage. (c) TLM curves and (d) $R_C$ and $R_{sheet}$ at $V_{gs} = 0\ V$ and $-40\ V$ (inset) as function of the temperature.

Similar measurements were performed as a function of the temperature, T, in the range 90 K to 400 K. Figure 3(c) shows that the linear behavior of the $R_T$ vs $d$ curves is preserved when the temperature is changed but their slope decreases with increasing T. Figure 3(d) reports the temperature dependence of $R_C$ and $R_{channel}$ evaluated at $V_{gs} = 0\ V$. $R_C$ remains constant over the 90-400 K temperature range while the graphene sheet resistance decreases, changing linearly from ∼ 22 $k\Omega/sq$ at 90 K to ∼ 12 $k\Omega/sq$ at 400 K with slope $dR_{sheet}/dT \sim -15\ \Omega/K$. The independence of $R_C$ on the temperature is confirmed also when $R_C$ is evaluated at $V_{gs} = -40\ V$, as shown in the inset of Figure 3(d). Conversely, a new feature appears in the temperature behavior of $R_{sheet}$ at $V_{gs} = -40\ V$: the sheet resistance decreases until the temperature reaches ∼ 200 $K$ and raises for $T \geq 200\ K$ up to ∼ 4.8 $\Omega/sq$ at $T = 400\ K$. Otherwise stated, a transition from a semiconducting to a metallic behavior occurs in graphene around $T \sim$

200 K, consistently with what has been observed before [89–92]. Similar TLM analyses have been conducted on devices of the same chip with graphene channel 2 $\mu m$ or 10 $\mu m$ wide or with different layout. The estimated contact resistance, normalized by the channel width, $R_C^* = R_C W$, and sheet resistances are summarized in Figure 4, showing a mean $R_C^*$ value of $\sim 2\ k\Omega \cdot \mu m$ and $R_{sheet} \sim 4\ k\Omega/sq$. We note that, owed to the zig-zag geometry, the TZ structure in Figure 4 (which is the previously analyzed one), shows a normalized contact resistance as low as $R_C^* \sim 700\ \Omega \cdot \mu m$ at $V_{gs} = -40\ V$ within the range of the good quality contacts typically reported in the literature [60,62–64]. The device-to-device sheet resistance fluctuations reported in Figure 4(b) can be attributed to local variations of the transferred graphene foil and different damage induced by the fabrication process.

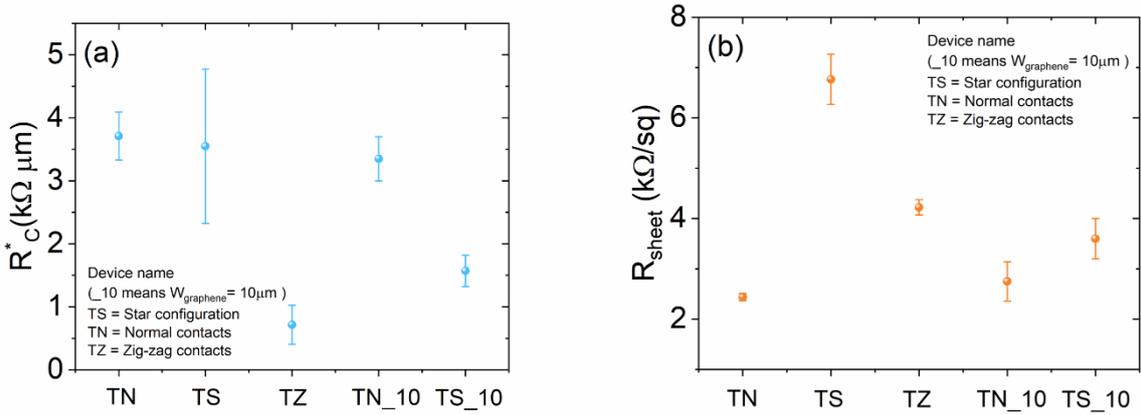

**Fig. 4.** (a) Contact resistance normalized by the channel width and (b) graphene sheet resistance calculated for different TLM devices (TS = star configuration, TN = straight contacts, TZ = contacts with zig-zag edges).

We further exploited the TLM measurements to study the dependence of the field-effect mobility, $\mu$, on the channel length $d$ and the temperature. The mobility was estimated from the slope $\frac{dI_{ds}}{dV_{gs}}$ of the linear part of the transfer characteristics as

$$\mu = \frac{d}{WC_{ox}V_{ds}} \frac{dI_{ds}}{dV_{gs}} \tag{8}$$

Figure 5(a) shows the hole mobility as function of channel length and displays a saturation for increasing channel length. Such a behavior can be expressed as $\mu = \mu_0\, d/(d + \lambda)$, where $\mu_0$ is the saturated mobility and $\lambda$ the mean free path [93]. From the fit of the experimental data, we obtain $\lambda \sim 1\ \mu m$ and $\mu_0 \sim 175\ cm^2 V^{-1} s^{-1}$. The saturation at channel length $d \gg \lambda$ corresponds to the establishment of a diffusive transport regime, while the mobility degradation at lower $d$ ($d \leq \lambda$) is an artifact due to the application of eq. (8) in a regime where the transport becomes quasi-ballistic or ballistic [93,94]. The influence of the temperature on the mobility, for the chosen device with $d = 110\ \mu m$, is shown in Figure 5(b), which indicates that most of the mobility degradation occurs for $T > 250\ K$ (~15% from its value at 90 K, $\mu_{90K} \sim 195\ cm^2 V^{-1} s^{-1}$). This behavior can be understood considering that phonon scattering in graphene becomes relevant only at higher temperatures [95–97].

The electron/hole mobilities evaluated from the numerical derivative of the $I_{ds} - V_{gs}$ curves according to eq. (8), are plotted in Figure 5(c) as function of $V_{gs} - V_{Dirac}$ (gate overdrive) for two channel lengths (1 and 10 $\mu m$, respectively). The mobilities show a minimum at the Dirac point, reach a maximum for increasing overdrive and decrease smoothly for $|V_{gs} - V_{Dirac}| > 10\,V$. The contact resistance, whose effect on the mobility is not eliminated in this type of analysis, could cause this decrease of $\mu$ with gate overdrive. To confirm such a hypothesis and obtain a more accurate $\mu$ vs $V_{gs}$ behavior, we considered the Y-function method as complementary approach to the TLM analysis.

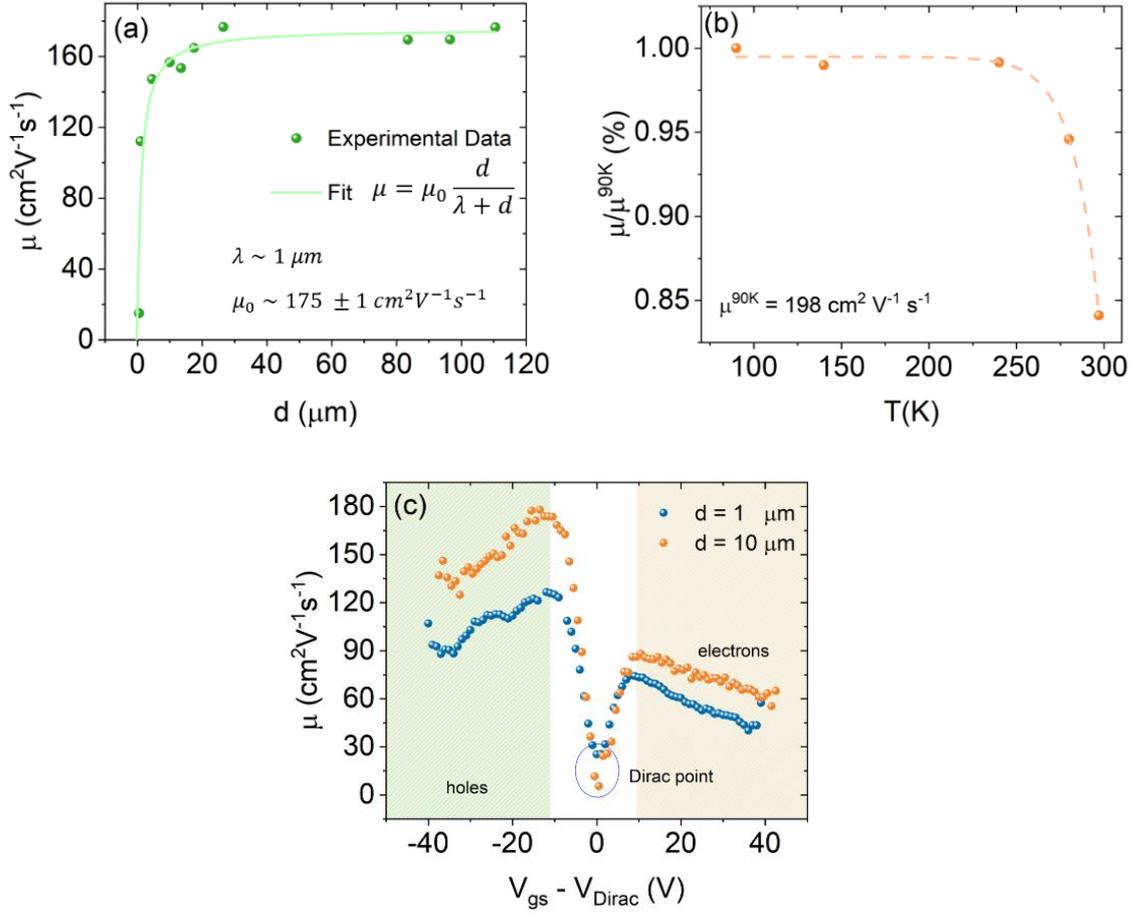

**Fig. 5**. Hole field-effect mobility plotted as function of channel length (a) and temperature (b). Panel (c) shows the hole and electron mobilities as function of $V_{gs} - V_{Dirac}$, at room temperature, for two channel lengths, 1 $\mu m$ (blue circles) and 10 $\mu m$ (blue circles).

The Y-function method has been successfully applied to eliminate the effects of the contact resistance on the estimation of the mobility and for the evaluation of $R_C$ itself. The mobility and contact resistance are obtained from the plots of $I_{ds}g_m^{-1/2}$ and $g_m^{-1/2}$ vs $(V_{gs}-V_{Dirac})$ as explained before (see eq. (5) and (6)). An example of these plots is shown in Figure 6 (a) and (b) for the p-branch of the transfer characteristic of the device with $d = 1\ \mu m$.

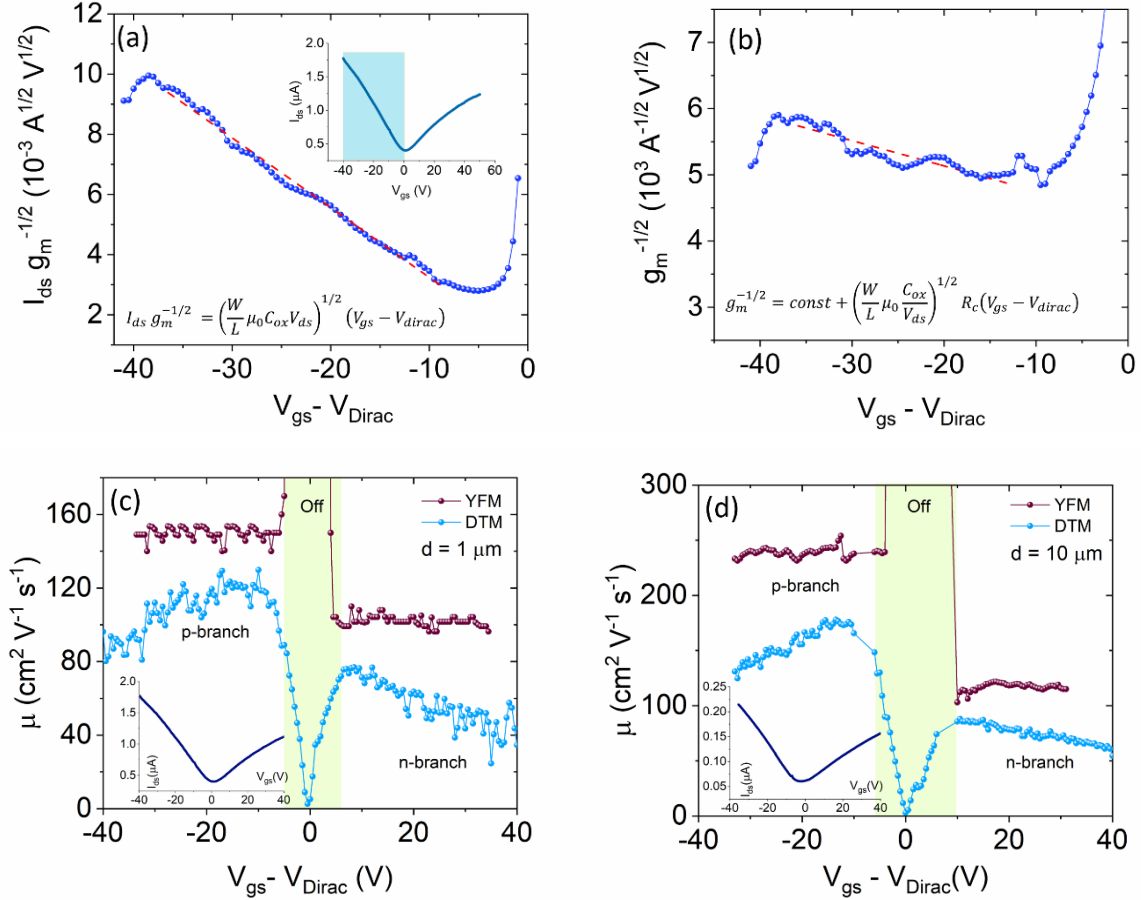

**Fig. 6.** (a) $I_{ds}/g_m^{1/2}$ and (b) $g_m^{-1/2}$ plotted as function of gate voltage using the Y-Function method. Comparison between $\mu$ vs $V_{gs} - V_{Dirac}$, extrapolated through the TLM method (light blue curves) and Y-function method (purple curves) for different the channel lengths (c) $1\ \mu m$ and (d) $10\ \mu m$.

From Figure 6(a), we obtained a field effect mobility $\mu \sim 160\ cm^2V^{-1}s^{-1}$, which used together with the data in Figure 6(b) yields $R_C = 310\ \Omega$ (or $R_C^* = 700\ \Omega \cdot \mu m$). These values are consistent with the results from the TLM analysis.

The plots of the mobility, with respect to $V_{gs}-V_{Dirac}$, are shown in Figures 6(c) and 6(d) for $1\ \mu m$ and $10\ \mu m$ channel lengths. The plots show $\mu$ unaffected by $R_C$ and confirm the independence of $\mu$ of the gate overdrive. As expected, the elimination of the effect of the contact resistance through the YFM removes the gate dependence of the field-effect mobility, thus confirming that it is only an artefact of the TLM method. We also note from Figures 6 (c) and (d) that removing $R_C$ results in significantly higher mobility, with over 80% increase.

## 5. Conclusions

In conclusion, we have fabricated and analyzed Ni-contacted graphene FETs and studied the gate and temperature dependence of the contact and channel resistance. We have measured devices with different geometrical structures and achieved competitive contact resistance using zig-zag shaped Ni contacts, also confirming the importance of contact geometry in the metal/graphene contact resistance.

We have found that the gate voltage modulates the contact and the channel resistance in a similar way but does not change the carrier mobility. We have also shown that the raising temperature decreases the carrier mobility, has a negligible effect on the contact resistance and can change the initial semiconducting behavior of the channel resistance into a metallic one, depending on the gate voltage. We have used two complementary methods, namely the TLM and the YFM, to show that, eliminating the detrimental effect of the contact resistance, can almost double the carrier field-effect mobility.